\documentclass[12pt]{article}
\usepackage{amsmath}
\usepackage{latexsym}
\usepackage{amssymb}

\numberwithin{equation}{section}
\newcommand{\be}{\begin{equation}}
\newcommand{\ee}{\end{equation}}
\newcommand{\beq}{\begin{equation}}
\newcommand{\eeq}{\end{equation}}
\newcommand{\ba}{\begin{eqnarray}}
\newcommand{\ea}{\end{eqnarray}}

\newcommand{\ber}{\begin{eqnarray}}
\newcommand{\eer}{\end{eqnarray}}

\newcommand{\beqar}{\begin{eqnarray}}
\newcommand{\eeqar}{\end{eqnarray}}

\newcommand{\dsl}
  {\kern.06em\hbox{\raise.15ex\hbox{$/$}\kern-.56em\hbox{$\partial$}}}
\newcommand{\sqg}{{\sqrt{-g}}}
\newcommand{\gumn}{{g^{\mu \nu}}}
\newcommand{\gdmn}{{g_{\mu \nu}}}
\newcommand{\Gdmn}{{G_{\mu \nu}}}
\newcommand{\goo}{{g^{00}}}
\newcommand{\gdoo}{{g_{00}}}
\newcommand{\eeqarr}{\end{eqnarray}}
\newcommand{\pa}{\partial}
\newcommand{\gab}{{g^{\alpha \beta}}}
\newcommand{\Roo}{{R_{00}}}
\newcommand{\Rrr}{{R_{rr}}}
\newcommand{\Rthet}{{R_{\theta \theta}}}
\newcommand{\nupp}{{\nu^{\prime\prime}}}
\newcommand{\nup}{{\nu^{\prime}}}
\newcommand{\labpp}{{\lambda^{\prime\prime}}}
\newcommand{\labp}{{\lambda^{\prime}}}
\newcommand{\Ttoo}{{\tilde{T}_{00}}}
\newcommand{\Too}{{T_{00}}}
\newcommand{\Tij}{{T_{ij}}}
\newcommand{\Ttij}{{\tilde{T}_{ij}}}
\newcommand{\Toi}{{T_{0i}}}
\newcommand{\Ttoi}{{\tilde{T}_{0i}}}
\newcommand{\Tmn}{{T_{\mu \nu}}}
\newcommand{\Ttmn}{{\tilde{T}_{\mu \nu}}}
\newcommand{\ZZ}{{\rm \kern 0.275em Z \kern -0.92em Z}\;}

\begin{document}

\title{Scalar Field for Breaking Lorentz and Diffeomorphism Invariance\footnote{Constrained Dynamics and Quantum Gravity, Cala Gonone, Sardinia, Italy, September 2005.}}
\author{R.Jackiw\\
{\small\itshape Center for Theoretical Physics}\\
{\small\itshape Department of Physics}\\
{\small\itshape Massachusetts Institute of
Technology} \\
{\small\itshape  Cambridge, Massachusetts 02139}}

\date{\small MIT-CTP-3423}

\maketitle

\thispagestyle{empty}
\begin{abstract}
A scalar field can be inserted in Maxwell and/or Einstein theory to effect symmetry breaking. Consequences of such a modification are discussed. Possible dynamics for the scalar field are presented.
\end{abstract}

\newpage
\section{Introduction}
Model field theories in which Lorentz symmetry and even CTP symmetry are violated have become the focus of attention for both theorists and experimentalists. The former build plausible models that incorporate violation of these sacred symmetries, the latter perform measurements that put ever-more stringent limits on the magnitude of symmetry violation.

Some years ago, my collaborators and I  produced the first such model in which electromagnetism is modified in a gauge invariant but Lorentz and CTP non-invariant fashion. The symmetry breaking interaction in the electromagnetic action arises from the electromagnetic Chern-Simons term (Gauss linking number). This 3-dimensional entity, when inserted into 4-dimensional Maxwell theory effects the symmetry violation through the dimensional mis-match. Specifically the 3-dimensional Chern-Simons term is inserted into the 4-dimensional theory with the help of an external 4-vector $v^\mu$, which picks out the space-time direction that determines how the Chern-Simons term is embedded in 4-dimensional space-time. We take this embedding vector to be time-like, $v^\mu = (\mu, {\bf 0})$ (so that spatial rotational symmetry is preserved). Alternatively, we view the embedding vector to be a gradient of a time dependent scalar $v_\mu = \partial_\mu \, \theta$. In either case an external quantity ($v^\mu \ \text{or}\ \theta$) is inserted into the theory. Astrophysical data may be used to limit the magnitude of the symmetry breaking modification: for all practical purposes we can conclude that  for electromagnetism Nature excludes this mechanism in its entirety \cite{rj3}, \cite{rj4}.

Exploring the formal similarities between a gauge theory (electromagnetism) and general relativity suggest a similar Chern-Simons modification of the gravity theory. However, physical results of the modified theory are quite different, since consequences of the (broken) diffeomorphism invariance overcome symmetry breaking effects, at least in the linear approximation.

In my talk I shall review the electromagnetic model, summarize the modified gravity theory, and discuss the space-times that arise when the embedding vector $v^\mu$ or scalar $\theta, v_\mu = \partial_\mu\, \theta$ are taken to be dynamical propagating quantites, rather than external, symmetry breaking parameters.

\section{Chern-Simons Modification of Maxwell Theory}

The  Chern-Simons term for an Abelian gauge theory on an Euclidean 3-space reads
\begin{equation}
CS (A) \equiv \frac{1}{4} \varepsilon^{ijk} F_{ij} A_k = \frac{1}{2} {\bf A} \cdot {\bf B}.
\label{1.1}
\end{equation}
The first expression is in tensor notation; the second in vector notation, with $\bf B$ being the magnetic field,
${\bf B} = {\bf \nabla} \times {\bf A}$. All indices are spatial $[i,j,k: x, y, z]$.
A related 4-dimensional formula in Minkowski space-time defines the topological Chern-Simons current
\begin{equation}
K^\mu = {^\ast F^{\mu \nu}} A_\nu,
\label{1.2}
\end{equation}
where ${^\ast F^{\mu \nu}}$ is the dual electromagnetic tensor.
\begin{equation}
{^\ast F^{\mu \nu}} = \frac{1}{2} \varepsilon^{\mu \nu \alpha \beta} F_{\alpha \beta}
\label{1.3}
\end{equation}
It is seen that the Chern-Simons term (\ref{1.1}) is proportional to the time $t\, (\mu=0)$ component of the Chern-Simons current, (\ref{1.2})
with the time dependence suppressed. Also the divergence of the topological current is the topological Pontryagin density.
\begin{equation}
\partial_\mu K^\mu = \partial_\mu ({^\ast F^{\mu \nu}} A_\nu) = \frac{1}{2} {^\ast F^{\mu \nu}} F_{\mu \nu}
\label{1.4}
\end{equation}

In Chern-Simons modified electromagnetism the Chern-Simons term (\ref{1.1}) (with field arguments extended to include $t$)
is added to the usual Maxwell Lagrangian.
\begin{equation}
I = \int d^4  x \left(- \frac{1}{4} \ F^{\mu \nu} \,F_{\mu \nu} + \frac{\mu}{2} \ \bf A \cdot \bf B \right)
\label{1.5}
\end{equation}
Here $\mu$, with dimension of mass, measures the strength of the extension. Formula (\ref{1.5}) may be alternatively presented in covariant
notation, with the help of an external, constant embedding 4-vector $v_\mu$.
\begin{eqnarray}
I= \int d^4  x \left(-\frac{1}{4} \ F^{\mu \nu} F_{\mu \nu} + \frac{1}{2} \  v_\mu {^*F}^{\mu \nu} A_\nu \right) \nonumber\\ [8pt]
v_\mu = (\mu, 0) \qquad \qquad \qquad \qquad \qquad
\label{1.6}
\end{eqnarray}
In spite of the presence of the vector potential, the action is gauge invariant: Under a gauge transformation it changes by a surface term,
since $\partial_\mu {^\ast F^{\mu \nu}} =0$. This can be made explicit by recognizing that in (\ref{1.6}) there occurs the Chern-Simons current
$ K^\mu$ (\ref{1.2}). Therefore, with the help of (\ref{1.4}) and an integration by parts the action acquires a gauge invariant form.
\begin{eqnarray}
I= \int d^4  x \left(-\frac{1}{4} \ F^{\mu \nu} F_{\mu \nu} + \frac{1}{4} \theta \ {^\ast \negthickspace F^{\mu \nu}} F_{\mu \nu} \right)
\nonumber\\ [8pt]
\partial_\mu \theta \equiv v_\mu \qquad \qquad \qquad \qquad \qquad
\label{1.7}
\end{eqnarray}
The external quantity is now $\theta$, which is taken as $\theta=\mu t$ so that (\ref{1.5}) and (\ref{1.6}) are reproduced.

Since the explicitly covariant formulations (\ref{1.6}), (\ref{1.7}) involve external, fixed quantities [a fixed constant embedding
vector $v_\mu$ in (\ref{1.6}); a fixed function $\theta$, linear in time, in (\ref{1.7})], we expect that Lorentz invariance is lost.
Also, since ${\bf A \cdot B}$, and ${^\ast F^{\mu \nu}} F_{\mu \nu}$ are axial quantities, parity is lost; but $C$ and $T$
are preserved, so CPT is also lost. To confirm these statements, we now look to the solutions of the modified equations of motion.

In the electromagnetic equations of motion, which follow from the Chern-Simons extended action, only Amp\`{e}re's law is modified.
\begin{equation}
-\frac{\partial {\bf E}}{\partial t}  + \nabla {\bf \times} \ {\bf B} = {\bf J} + \mu {\bf B}
\label{1.8}
\end{equation}
All other Maxwell equations continue to hold. Also the consistency condition on  (\ref{1.8}) remains as in Maxwell theory: 
the charge density $\rho = {\bf \nabla} \cdot {\bf E}$ and the current $\bf J$ must satisfy their continuity equation, as is seen by taking
the divergence of (\ref{1.8}) and using ${\bf \nabla \cdot B}=0$.

The modification that we have constructed is particularly felicitous for the following reasons.
\begin{description}
\item{(i)}
Gauge invariance is maintained, so the photon continues to possess just two independent polarizations.
\item{(ii)}
Eg.(\ref{1.8}) is not a radical departure; it has played previous roles in physical theory: in plasma physics one frequently
replaces the source current $\bf J$ with a magnetic field $\bf B$. Of course, we are not working with a collective/phenomenological theory, like
plasma physics, rather we are examining the  feasibility of (\ref{1.8}) for fundamental physics.
\end{description}

To assess the actual physical content of the Chern-Simons extended electromagnetism, and its associated symmetry breaking,
we have examined some solutions. We found that in the source-free case, plane waves continue to solve the extended equations.
The photon posseses two independent polarizations, (as anticipated from gauge invariance) however they travel at velocities which
differ from the velocity of light (thus Lorentz boost invariance is lost---as anticipated) and also the two polarizations
travel with velocities that differ from each other (thus parity invariance is lost--as anticipated).

The fact that the two photon helicities travel (in vacuum) with different velocities makes empty space behave as a birefringent medium.
Consequently linearly polarized light, passing through this birefringent environment, undergoes a Faraday-like rotation, which can be looked
for in observations of light from distant galaxies. Much data exists on this phenomenon, and the conclusion is unavoidable: there is no such
effect in Nature;
$\mu=0$ is required. This was asserted in our initial investigation \cite{rj3}, and the many other analyses carried out in the
intervening years support that conclusion (see e.g. \cite{rj4}).

\section{Chern-Simons Modification of Einstein Theory}
{\bf A. Gravitational Chern-Simons term in 3-space}.\\ 
The 3-dimensional, gravitational Chern-Simons term can be presented in terms
of the 3-dimensional Christoffel connection $^3\Gamma_{i q}^p$, \cite{rj2}
\begin{equation}
CS(\Gamma) =   \varepsilon^{ijk} \ (\frac{1}{2}
\ ^3\Gamma^p_{i q} \, \partial_j  {^3 \Gamma^q_{k p}} + \frac{1}{3}  {^3\Gamma}^p_{iq} \, {^3\Gamma}^q_{j r} \, ^3\Gamma^r_{kp}),
\label{2.1}
\end{equation}
but it is understood that the Christoffel connection takes the usual expression in terms of the metric tensor, which is the fundamental
variable. Variation with respect to the metric tensor of the intergrated Chern-Simons term results in the 3-dimensional ``Cotton tensor";
which involves a covariant curl of the 3-dimensional Ricci tensor $^3R^i_j$.
\begin{equation}
\frac{\delta}{\delta g_{ij}} \int d^3 x CS(\Gamma) =-\sqrt{g} \ {^3C^{ij}} = \frac{1}{2}
\varepsilon^{imn} \ {^3D_m} {^3R^j_n} + i\leftrightarrow j
\label{2.2}  
\end{equation}
${^3C^{ij}}$ is symmetric, traceless and covariantly conserved. It vanishes if and only if the 3-dimensional metric tensor is conformally flat.
A related formula gives the 4-dimensional Chern-Simons current $K^\mu$,
\begin{equation}
K^\mu = 2 \varepsilon^{\mu}  {^{\alpha \beta \gamma}} \left [\frac{1}{2} ~ \Gamma^\sigma_{\alpha \tau} \, \partial_ \beta
~ \Gamma^\tau_{\gamma \sigma} + \frac{1}{3} ~  \Gamma^\sigma_{\alpha \tau} \,  \Gamma^\tau_{\beta \eta} \Gamma^\eta_{\gamma \sigma} \right],
\label{2.3}
\end{equation}
whose divergence is the topological Pontryagin density.
\begin{equation}
\partial_\mu K^\mu = \frac{1}{2} {^*R}^\sigma_{~\tau} \ {^{\mu \nu}} \ R^\tau {_{\sigma \mu \nu}} \equiv \frac{1}{2} {^*RR}
\label{2.4}
\end{equation}
Here $R^\tau_{~\sigma \mu \nu}$ is the Riemann curvature tensor and $^\ast R^\sigma_{~ \tau} {^{\mu \nu}}$ is its dual.
\begin{equation}
^*R^{\sigma \ \mu \nu}_{\, \,\tau} = \frac{1}{2} ~ \varepsilon^{\mu \nu \alpha \beta} R^\sigma_{~{\tau \alpha \beta}}
\label{2.5}
\end{equation}
[Notation: $(i, j,...)$ are 3-dimensional, spatial indices, and 3-dimensional geometric entities are decorated with the superscript ``3". Undecorated
geometric entities are 4-dimensional, and Greek indices label the 4 space-time coordinates.]
Note that unlike in the vector case, the Chern-Simons term (\ref{2.1}) is not the time component $K^0$, because the former contains
3-dimensional Christoffel entities, while 4-dimensional ones are present in $K^0$. This variety allows various
extensions general relativity.\\ \vspace{-6pt}

{\noindent {\bf B. Gravitational Chern-Simons term in 4-space.}}

In analogy with the electromagnetic formulation (\ref{1.7}), we choose to extend Einstein theory by adopting the
action \cite{rj6}
\begin{eqnarray}
I&=& \frac{1}{16 \pi G} \int d^4 x \left( \sqrt{-g} R + \frac{1}{4} \theta {^*RR} \right) \nonumber \\ [8pt]
&=& \frac{1}{16 \pi G} \int d^4 x \left(\sqrt{-g} R - \frac{1}{2} v_\mu K^\mu \right), \qquad v_\mu \equiv \partial_\mu \theta.
\label{2.6}
\end{eqnarray}
The first contribution is the usual Einstein-Hilbert term involving the Ricci scalar $R$. The modification involves an external
quantity: $\theta$ in the first equality; $\partial_\mu \theta \equiv v_\mu$ in the second equality, which follows from the first by (\ref{2.4})
and an integration by parts. Eventually we shall take the embedding vector $v_\mu$ to possess only a time component, and $\theta$ to depend
solely on time. So then our modification (\ref{2.6}) involves the time component of 4-dimensional Chern-Simons current (\ref{2.3}) 
[rather than the 3-dimensional Chern-Simons term (\ref{2.1})].

The equation of motion that emerges when (\ref{2.6}) is varied with respect to $g_{\mu \nu}$ is
\begin{equation}
G^{\mu \nu} + C^{\mu \nu} = -8 \pi G T^{\mu \nu}.
\label{2.7}
\end{equation}
Here $G^{\mu \nu}$ is the covariantly conserved (Bianchi identity) Einstein tensor, $G^{\mu \nu} = R^{\mu \nu} -\frac{1}{2} g^{\mu \nu} R$.
We have inserted a source with strength $G$ (Newtons constant) consisting of the matter energy-momentum tensor
$T^{\mu \nu}$, which also is convariantly conserved, since we assume matter to be conventionally, covariantly coupled to gravity.
$C^{\mu \nu}$ is the term with which we are extending the Einstein theory.
\begin{equation}
\sqrt{-g} \ C^{\mu \nu} = \frac{\delta}{\delta g_{\mu \nu}} \ \frac{1}{4} \int d^4 x \ \theta
{^\ast RR} = -\frac{1}{2} \left( v_\sigma \varepsilon^{\sigma \mu \alpha \beta} D_\alpha
R^\nu_\beta +  v_{\sigma \tau} {^\ast R^{\tau \mu \sigma \nu}} + \mu \leftrightarrow \nu \right)
\label{2.8}
\end{equation}
$C^{\mu \nu}$ is manifestly symmetric; it is traceless because $^*RR$ is conformally invariant.
$C^{\mu \nu}$'s first term (involving the curl of $R^\nu_\beta$) is similar to the 
3-dimensional $\sqrt{g} {^3C^{i j}}$ (\ref{2.2}). Even the second term can be viewed as a generalization
from 3-dimensions: it involves only the Weyl tensor part of Riemann tensor, which vanishes in 3 dimensions.
[Cotton defined his tensor in arbitrary dimensions $d$, and his definition is equivalent to ours in $d=3$, where it is also given by the variation of
the 3-d gravitational Chern-Simons term, as is (\ref{2.2}). \cite{rj7} However for $d\ne 3$, Cotton's tensor does not appear to have a variational
definition. Our $d=4$ Cotton-like tensor in (\ref{2.8}) does possess a variational definition, at the the expense of introducing non-geometrical
entities like
$\theta \ \mbox{and}\ v_\mu$.]

Finally we must examine $D_\mu C^{\mu \nu}$, whose vanishing is a consistency requirement on (\ref{2.7}).
However, an explicit evaluation (which involves only geometric identities) shows that, unlike in 3 dimensions, $C^{\mu \nu}$ is not covariantly
conserved. Rather
\begin{equation}
D_\mu C^{\mu \nu} = \frac{1}{8 \sqrt{-g}} \ v^\nu {^*RR}.
\label{2.9}
\end{equation}
Thus the vanishing of $^\ast RR$ is a consistency condition of the new dynamics: every solution to (\ref{2.7})
 will necessarily lead to vanishing Pontryagin density.
\begin{equation}
G^{\mu \nu} + C^{\mu \nu} = -8 \pi G T^{\mu \nu} \Rightarrow {^\ast RR} = 0
\label{2.10}
\end{equation}

We may derive and understand the expression for the covariant divergence of $C^{\mu \nu}$ by examining
the response of our addition to changes in the coordinates. With the infinitesimal transformation
\begin{equation}
\delta x^\mu = -f^\mu (x)
\label{2.11}
\end{equation}
we have
\begin{equation}
\delta g_{\mu \nu} = D_\mu f_\nu + D_\nu f_\mu.
\label{2.12}
\end{equation}
The Hilbert Einstein action is of course invariant. To assess the variance properties of our modification,
we can proceed in two ways. First observe that ${^\ast RR}$ is scalar density, so it transforms as
$\delta {(^\ast RR)} = \partial_\mu (f^\mu {^\ast RR})$. $\theta$ is an external quantity, therefore we do not transform
it.
\begin{subequations}
\begin{eqnarray}
\delta I_{CS} = \frac{1}{4} \int d^4 x \ \theta \delta (^\ast RR)&=& \frac{1}{4} \int d^4 x
\theta \partial_\mu (f^\mu {^\ast RR}) \nonumber\\ [8pt] &=&-\frac{1}{4} \int d^4 x
v_\mu f^\mu {^\ast RR}
\label{2.13a}
\end{eqnarray}
Alternatively, we may vary $I_{CS}$, by varying $g_{\mu \nu}$ according to (\ref{2.12}) , and using the definition
(\ref{2.8}) for $C^{\mu \nu}$.
\begin{eqnarray}
\delta I_{CS}&=& \int d^4 y\  \frac{\delta}{\delta g_{\mu \nu} (y)} \ \left(\frac{1}{4} \int d^4 x \ \theta \ {^\ast RR} \right)
2 D_\mu f_\nu (y) \nonumber\\ [8pt]
&=& 2 \int d^4 x \ \sqrt{-g} \ C^{\mu \nu} D_\mu f_\mu = -2 \int d^4 x \ \sqrt{-g}
\ (D_\mu C^{\mu \nu} ) f_\nu 
\label{2.13b}
\end{eqnarray}
\end{subequations}
Equating the two expressions for $\delta I_{CS}$ establishes (\ref{2.9}), and also demonstrates that $^\ast RR$ is a measure of
the failure of diffeomorphism invariance. But $^\ast RR$ vanishes as a consequence of the equation of motion, so
is some sense diffeomorphism invariance is dynamically reinstated.

For another perspective, consider a variant of our model, where $\theta$ in (\ref{2.6}) is  a dynamical
variable, not an externally fixed quantity. There is no kinetic term; $\theta$ acts as a Lagrange multiplier and covariance is maintained. This is because we postulate that under diffeomorphisms (\ref{2.11})
$\theta$ transforms as a scalar,
\begin{equation}
\delta \theta = f^\mu \partial_\mu \theta = f^\mu v_\mu,
\label{2.14}
\end{equation}
and (\ref{2.13a}) acquires the additional contribution
\begin{equation}
\frac{1}{4} \int d^4 x \, \delta \theta ~({^\ast RR}) = \frac{1}{4} \int d^4 x \ v_\mu \ f^\mu
\ {^\ast \negthickspace RR},
\label{2.15}
\end{equation}
which cancels (\ref{2.13b}), showing that the Chern-Simons modification with dynamical $\theta$ is
diffeomorphism invariant. Now let us look at the equations of motion in this variant of modified
gravity: varying $g_{\mu \nu}$ still produces (\ref{2.7}); varying $\theta$, now acting as a Lagrange
multiplier, forces $^\ast RR$ to vanish, but that requirement is already implied by (\ref{2.7}), (\ref{2.10}).
Thus the equations of the fully dynamical, and diffeomorphism invariant theory coincide with the equations
of the non-invariant theory, where $\theta$ is a fixed, external quantity.

Formula (\ref{2.13a}) shows that when $v_\mu$ is chosen to have only a time component, $v_\mu
=(\frac{1}{\mu}, \bf 0)$; equivalently $\theta = t/ \mu$, then $I_{CS}$ is invariant under all space-time
reparametrizations of the spatial coordinates, and also of shifts in time: $f^0 = \mbox{constant},
f^i \mbox{arbitrary}$. Henceforth we make this choice for $v_\mu\ \mbox{and} \ \theta$.\\ 

{\noindent \bf C. Physical effects of the Chern-Simons term in 4-d gravity.}\\
We examine some physical processes in the Chern-Simon modified gravity theory.
\begin{description}
\item{(i)} 
It is important that the  Schwarzschild solution continues to hold; thus our theory  passes the 
``classic" test of general relativity. The result is established in two steps.
First we posit a stationary form for the metric tensor
\begin{equation}
g_{\mu \nu} = \left(
\begin{array}{cc}
N& 0\\
0& g_{ij}
\end{array} \right),
\label{2.16}
\end{equation}
with time-independent entries. It follows that $C^{00}\ \mbox{and}\ C^{n0}= C^{0n}$ vanish. Also one finds
that the spatial components of $C^{\mu\nu}$ reproduce the 3-dimensional Cotton tensor.
\begin{equation}
\sqrt{-g} \ C^{ij}  =  \sqrt{g}^3 \ C^{ij}
\end{equation}
Next, we make the spherically symmetric {\it Ansatz}, and find that $C^{ij}$ vanishes. Evidently also
$^\ast RR$ must vanish on the Schwarzschild geometry, because the modified equations are satisfied.
Since the Kerr geometry, carries non vanishing $^\ast RR$, it will not be a solution to the extended
equations. It remains an interesting, open question which deformation of the Kerr geometry satisfies
the Chern-Simons modified equations. 
\vspace{8pt}

\item{(ii)} 
Next we perform a linear analysis by expanding the metric tensor around a flat background
$g_{\mu \nu} = \eta_{\mu \nu} + h_{\mu \nu}$. The purpose of the linear analysis is to determine the propagating degrees of freedom, to study
the nature of small disturbances (gravity waves) and to illuminate the construction of an energy-momentum (pseudo)
tensor, which is symmetric and divergence-free.

Keeping only the linear portions of the Einstein tensor and $C_{\mu \nu}$, we
verify that both $G^{linear}_{\mu \nu} \ \mbox{and}\ C^{linear}_{\mu \nu}$ are divergence-free.
\begin{equation}
\partial^\mu G^{linear}_{\mu \nu} = 0 = \partial^\mu C^{linear}_{\mu \nu}
\label{2.17}
\end{equation}
This is seen from the explicit formulas. It also follows from the observation that the exact equation
$D^\mu G_{\mu \nu} =0$ implies the above result for $G^{linear}_{\mu \nu}$; moreover, from (\ref{2.9})
we see that $D^\mu C_{\mu \nu}$ is of quadratic order, hence the above result for $C^{linear}_{\mu \nu}$
holds also.
It is further seen that the linear portions are invariant under the ``gauge" transformation
\begin{equation}
h_{\mu \nu} \to h_{\mu \nu} + \partial_\mu \lambda_\nu + \partial_\nu \lambda_\mu
\label{2.18}
\end{equation}
\end{description}

In the Einstein theory, one decomposes $h_{\mu \nu}$ into temporal parts, and purely spatial parts $h^{ij}$.
The latter is further decomposed into its trace, its longitudinal part, and its traceless transverse part,
denoted by $h^{ij}_{TT}$. One then finds from the linear equations that, with the exception
of $h^{ij}_{TT}$, all other components of $h_{\mu \nu}$ are either non-propagating or can be eliminated
by the gauge transformation (\ref{2.18}). Only $h^{ij}_{TT}$ survives and it is governed by a 
d'Alembertian. Since a symmetric, transverse and traceless $3\times 3$ matrix possesses
two independent components, one concludes that in Einstein's theory small gravitational disturbances
are waves, with two polarizations, each moving with the velocity of light (governed by the d'Alembertian).

None of this changes where $C^{linear}_{\mu \nu}$ is included. Again only $h^{ij}_{TT}$ propagates,
governed by the d'Alembertian. Explicitly the modified equation for $h^{ij}_{TT}$ reads
\begin{eqnarray}
(\delta^{im} \delta^{jn} + \frac{1}{2\mu} \varepsilon^{ipm} \ \delta^{nj} \ \partial_p + \frac{1}{2\mu}
\varepsilon^{jpm} \ \delta^{ni} \ \partial_p) \ \Box \ h^{mn}_{TT} \nonumber\\
=-16 \pi \ G \ T^{ij}_{TT}. \qquad \qquad \qquad \qquad \qquad \qquad
\label{2.19}
\end{eqnarray}
$T^{ij}_{TT}$ is the transverse traceless part of the stress tensor. The new terms are the $(\mu^{-1})$ contributions; they
involve only spatial derivatives. One may consider that the left side of (\ref{2.19}) involves an operator
acting on $\Box \ h^{mn}_{TT}$.
\begin{subequations}\label{2.20}
\begin{equation}
\mathcal{O}^{ij}_{~ \ mn} \Box \ h^{mn}_{TT} = -16 \pi G \ T^{ij}_{TT}
\label{2.20a}
\end{equation}
Acting on this equation with the inverse operator $\mathcal{P} = \mathcal{O}^{-1}$ shows that the effect
of the entire extension is to modify the source
\begin{eqnarray}
\Box \ h^{mn}_{TT}&=&-16 \pi G \ \mathcal{P}^{mn}_{~\ \ \ ij} \ T^{ij}_{TT} \nonumber\\
&\equiv&-16 \pi G \ \tilde{T}^{ij}_{TT} 
\label{2.20b}                                                                 
\end{eqnarray}
\end{subequations}

Thus we see that in sharp contrast to the electromagnetic case, the Chern-Simons modification of gravity
does not change the velocity of gravity waves and there is no Faraday rotation. It is also noteworthy
that the reduction to 2 degrees of freedom (2 polarizations) takes place also in the extended theory.
Such a reduction of degrees of freedom is considered to be a consequence of gauge invariance, here diffeomorphism
invariance, which evidently continues to hold on our modified theory.

There does exist a physical manifestation of the extension. Although the velocities of the two polarizations
are the same, their intensities differ, due to the modification of the source $(T^{ij}_{T T} \to \tilde{T}^{ij}_{TT})$.
One finds for  a weak modification (large $\mu$) that the ratio  of the intensity of waves with
negative helicity to those with positive helicity is
\begin{equation}
\frac{-}{+} = \left(1+ \frac{4\omega}{\mu}\right),
\end{equation}
where $\omega$ is the frequency. This puts into evidence the parity violation of the modification.

Finally we turn to the topic of the energy-momentum (pseudo) tensor. A straight forward approach
to this problem in the Einstein theory is to rewrite the equation of motion by decomposing the Einstein
tensor $G_{\mu \nu}$ into its linear and non-linear parts, and moving the non-linear terms to the ``right"
side, summing it with the matter energy-momentum tensor.
\begin{equation}
G^{linear}_{\mu \nu} = -8\pi \ G \left(T_{\mu \nu} + \frac{1}{8\pi  G} \ G^{non-linear}_{\mu
\nu}\right)
\end{equation}
Clearly $G^{linear}_{\mu \nu}$ is symmetric and conserved, therefore, so must be the right side, which is now renamed
as total (gravity + matter) energy-momentum (pseudo) tensor.
\begin{equation}
\tau_{\mu \nu} = T_{\mu \nu} + \frac{1}{8\pi  G} \ G^{non-linear}_{\mu \nu}
\end{equation}

Exactly the same procedure works in the extended theory. We present the equation of motion (\ref{2.7}) as
\begin{equation}
G^{linear}_{\mu \nu} + C^{linear}_{\mu \nu} = -8 \pi \thinspace G \, \left(T_{\mu \nu} + \frac{1}{8 \pi G} (G^{non-linear}_{\mu \nu} +
C^{non-linear}_{\mu \nu}) \right).
\end{equation}
We have already remarked that the left side is divergenceless. Thus we can identify a symmetric and conserved energy-
momentum (pseudo) tensor as
\begin{equation}
\tau_{\mu \nu} = T_{\mu \nu} + \frac{1}{8\pi G} (G^{non-linear}_{\mu \nu} + C^{non-linear}_{\mu \nu}).
\label{2.26}
\end{equation}
It is striking that this structure is present in a theory that seems to violate Lorentz invariance!

In Ref. \cite{rj8} there is a survey of other gravitational energy-momentum (pseudo) tensors for Einstein's theory that
differ from each other by super potentials. In particular there is described a Noether construction with a Belinfante
improvement, which also yields a symmetric, conserved energy-momentum (pseudo) tensor tied to the Poincar\'{e} invariance of the Einstein
theory. It would be interesting to reconsider this construction in the extended theory and to compare the result to (\ref{2.26}).\\

{\noindent \bf D. Space-Time produced by a time-dependent scalar field}\\

The $\theta$ variable, which we introduced into the gravity theory, does not possess independent dynamics. Whether it is a prescribed external quantity or a Lagrange multiplier, it leads to the vanishing of the Pontryagin density.

In the literature there are various dynamical models which give rise to the $\theta$ or  $\partial_\mu\, \theta \equiv v_\mu$ variables through self-consistent equations. However, a non-vanishing value requires exotic or unatural forms for the $\theta$ or $v^\mu$ Lagrangians. So we have attempted a simpler approach based on conventional dynamics: gravity with a minimally coupled $\theta$ field \cite{tenpi}. Indeed the $\theta {^*\! RR}$ term need not be considered, because the geometries that we find solving our equations produces a vanishing $\theta {^*\! RR}$ and Cotton-like tensor (\ref{2.8}).

 The field equations are Einstein equation  and the equation of motion for $\theta$,
\ba
\Gdmn &=& 8\pi\, G\ \Tmn \label{eq:1}\\
D^2 \theta &=& \tfrac{1}{\sqg}\ \pa_\mu\ (\sqg\ \gumn\, \pa_\nu \, \theta) = 0 \label{eq:2}
\ea 
where  the energy-momentum tensor $\Tmn$ given by
\be
\Tmn = \pa_\mu \, \theta \pa_\nu \, \theta - \tfrac{1}{2}\ \gdmn\, \gab\, \pa_\alpha \, \theta\, \pa_\beta\, \theta. \label{eq:3}
\ee
Using (\ref{eq:3}), the Einstein equation may be written as 
\be
R_{\mu\nu} = 8 \pi\, G \ \pa_\mu \, \theta\, \pa_\nu\, \theta. 
\label{eq:4}
\ee

We considered two kinds of solutions: (i) a spherically symmetric, time-dependent metric, which we call a ``vacuum" configuration; (ii) a Robertson-Walker metric, which may be called a ``cosmological" solution. For both cases, only the diagonal components of $R_{\mu\nu}$ are non-vanishing. The space-time component for Einstein equation (\ref{eq:4}),
\be
R_{0 i} = 0 = 8\pi\, G\, \dot{\theta}\, \pa_i\, \theta ,
\label{eq:5}
\ee
requires either $\dot{\theta} = 0$ or  $\pa_i\, \theta= 0$. We posit the latter eventuality so that $\theta$ depends only on time. Then the remaining non-diagonal components $R_{ij}, i \ne j$, lead to vacuous equations. The diagonal components of $R_{\mu\nu}$ provide three differential equations:
\begin{subequations}\label{eq:6}
\ba
\Roo &=& 8\pi\, G \, \dot{\theta}^2 \label{eq:6a}\\
\Rrr &=& 0  \label{eq:6b}\\
R_{\theta \theta} &=& 0. \label{eq:6c}
\ea
\end{subequations}
$R_{\varphi \varphi} = \sin^2 \theta\, \Rthet$ does not provide a new equation. For $\theta$, which depends only on $t$, the equation of motion (\ref{eq:3}) becomes 
\be
\pa_0 (\sqg\ \goo \, \dot{\theta}) = 0.
\label{eq:7}
\ee
The above four equations, (\ref{eq:6a}, \ref{eq:6b}, \ref{eq:6c}) and (\ref{eq:7}) are key equations for both the vacuum and cosmological solutions. [Note that in Eq. \eqref{eq:6c} and at the beginning of the first line below \eqref{eq:6c} $\theta$ is the polar angle, not to be confused with the scalar field, which is denoted by $\theta$ throughout the paper.]

\subsection*{Vacuum Solution}
The most general form of the line element of a spherically symmetric, time-independent metric may be parameterized as
\be
d s^2 = e^\nu\, d t^2 - e^\lambda\, d r^2 - r^2\, d \Omega^2,
\label{eq:8}
\ee
with $\lambda$ and $\nu$ functions of only $r$. For this metric, the solution to (\ref{eq:7}) is given by
\be
\theta (t) = t/\mu,
\label{eq:9}
\ee
where $\mu$ is an arbitrary constant. Then equations (\ref{eq:6a})-(\ref{eq:6c}) lead to the following differential equations.
\begin{subequations}\label{eq:10}
\ba
\tfrac{\nup}{r} + \tfrac{1}{2}\ (\nupp +\tfrac{1}{2}\ \nup^2 - \tfrac{1}{2} \labp \nup ) &=& 8\pi \, G\, e^{(\lambda - \nu)}/\mu^2 \label{eq:10a}\\
\tfrac{\labp}{r} - \tfrac{1}{2}\ (\nupp + \tfrac{1}{2}\ \nup^2 - \tfrac{1}{2}\ \labp\, \nup ) &=& 0 \label{eq:10b}\\
\nup - \labp = \tfrac{2}{r} \ (e^\lambda - 1) \label{eq:10c} \hspace{.5in}
\ea
\end{subequations}
(Prime denotes differentiation with respect to $r$.) After expressing $\nup$ in terms of $\lambda$ by use of (\ref{eq:10c}), (\ref{eq:10b}) may be written as
\be
\labpp + \tfrac{3\labp}{r}\ (e^\lambda - 1) + \tfrac{2}{r^2}\ (e^\lambda - 1) (e^\lambda - 2) = 0,
\label{eq:11}
\ee
while the sum of (\ref{eq:10a}) and (\ref{eq:10b}) becomes
\be
e^{-\nu} = \tfrac{2}{m^2r^2}\ [1+(r\labp - 1) \, e^{-\lambda}],
\label{eq:12}
\ee
where $m^2 \equiv 8\pi\ G/\mu^2$. [Because Eq. (\ref{eq:11}) is derived by differentiating (\ref{eq:10c}), it possesses spurious solutions, which do not satisfiy (\ref{eq:10a} -- \ref{eq:10c}). An example is $e^\lambda = r/(r-c)$. This solves (\ref{eq:11}), and (\ref{eq:10c}) leads to vanishing left side of (\ref{eq:10a}); {\it viz.} it is the Schwarschild solution, which requires the right side (\ref{eq:10a}) to vanish. Substituted in (\ref{eq:12}) the spurious solution gives $e^{-\lambda} = 0$.]. There are two self-evident solutions of (\ref{eq:11}): $e^\lambda =1$ and $e^\lambda =2$.  However, $e^\lambda = 1$ is a special case of the spurious solution with $c = 0$. The solution $e^\lambda = 2$ gives $e^\nu = m^2 r^2$ leading to the Wyman line-element \cite{pi02}
\be
d s^2 = m^2\, r^2\, d t^2 - 2 d r^2\, - r^2\, d \Omega^2.
\label{eq:13}
\ee

It appears that all solutions tend to the above expression at large distances. This follows from an analysis in which it is assumed that the asymptotic expression can be expanded in dominant and subdominant terms. We find,
\begin{subequations}
\ba
\lambda &=& l n\, 2 + \tfrac{\alpha}{m r} \ \cos\ (\sqrt{3}\ l n\, m r\, + \beta) + ...\label{eq:14a}\\
\nu &=& 2 \ l n \, m r - \tfrac{\alpha}{m r}\  [\cos\ (\sqrt{3} \ l n\, m  r + \beta) + \sqrt{3} \ \sin\ (\sqrt{3}\ l n \, m r + \beta) ] \label{eq:14b}
\ea
\end{subequations}
where $\alpha$ and $\beta$ are arbitrary parameters. The next sub-leading terms do not introduce any additional parameters and behave as $(m r)^{-2}$ times trigonometric functions with twice the above arguments.

Thus a scalar field, linear in time, produces self-consistently a 2-parameter family of static, rotationally symmetric space-times, none of which is asymptotically flat. The only analytic solution we found, {\it i.e.} the expression in (\ref{eq:13}), produces the unique special case in which there are no arbitrary parameters.

The system possesses scale invariance, $\lambda(r) \to \lambda(c r)$. The explicit Weyman vacuum solution is the unique solution which is scale invariant; other solutions are scale covariant, in the sense that scale transformations change the two arbitrary parameters ($\alpha, \beta$) of the solutions. 

It is not apparent that the space-time described by the line-element (\ref{eq:13}) can be employed as a physically acceptable background about which gravity theory should be expanded. It possesses a singularity at $ r= 0$, which acts as an attractor for geodesics, whose paths can be determined as usual from the geodesic equation. On the plane where the polar angle is $  \pi /2$, the path has a simple form,
\ba
r (t) &=& \tfrac{r_0}{\cosh \omega t} \nonumber\\
\phi &=& \sqrt{2}\ \bar{\omega} t \label{eq:15}
\ea
where $\omega, \bar{\omega}$ and $r_0$ are constants of motion satisfying $\omega^2 +\bar{\omega}^{2} = \frac{m^2}{2}$. A particle starting out at $r_0$ spirals into the origin in infinite time.

\subsection*{Cosmological Solution}
For the Robertson-Walker metric,
\be
d s^2 = d t^2 - a^2(t)\ \big[ \tfrac{d r^2}{1- k r^2}\ + r^2 d \, \Omega^2\big] 
\label{eq:16}
\ee
where $k = \pm , 0$, the solution to (\ref{eq:7}) satisfies
\be
\dot{\theta} (t) =  a(t)^{-3} /\mu 
\label{eq:17}
\ee
where $\mu$ is an arbitrary constant. The Einstein equations (\ref{eq:6}) read
\begin{subequations}\label{eq:18}
\ba
- 3 \tfrac{\ddot{a}}{a} = 8 \pi\, G \ \mu^{-2}\, a^{-6} \equiv m^2\, a^{-6} \label{eq:18a}\\
(a \ddot{a} + 2\dot{a}^2 + 2 k) \ \hat{g}_{ij} = 0
\label{eq:18b}
\ea
\end{subequations}
where $\hat{g}_{ij}$ is the metric for 3-dimensional comoving coordinates. The first integral of (\ref{eq:18a}) becomes
\be
\dot{a}^2 = \tfrac{1}{6}\ m^2\, a^{-4} - c
\label{eq:19}
\ee
where $c$ is an integration constant. Using (\ref{eq:18a}) and (\ref{eq:19}), one finds that $c=k$. The final integration of (\ref{eq:19}) involves elliptic functions for $k\ne 0$, so for simplicity we consider a flat Robertson-Walker metric, $k=0$.

At $k=0$ we find that
\be
a(t) = (\tfrac{3}{2})^{1/6}\, (m t)^{1/3}.
\label{eq:20}
\ee
Substituting (\ref{eq:20}) into (\ref{eq:17}), yields
\be
\theta (t) = \tfrac{1}{\sqrt{12\pi\, G}}\ ln\, m t.
\label{eq:21}
\ee
For this scenario, the universe expands as $t^{1/3}$, which is different from the expansion due to radiation or matter domination in the standard cosmology. It will be interesting to determine the parity violating effects of our $\theta {^*\! RR}$ extension on the cosmic microwave background radiation.

The geodesic motion is described by
\be
{\bf r} (t) = {\bf r}_0 \sqrt{\frac{(12\pi G)^{1/3}}{9}\ r^2_0 + t^{2/3}}
\label{eq:348}
\ee
where ${\bf r}_0$ is constant.
\subsection*{Hydrodynamic Formulation}
More information about the space-times produced by a homogenous, but time-dependent scalar field can be obtained from the hydrodynamic formulation of our equations. It is known that the energy-momentum tensor of a scalar field that depends only on time has an ideal fluid representation, provided $g_{0i}$ vanishes \cite{pi03}. If $g_{0i} = 0$, it follows that $\goo = 1/g_{00}$. For both of our scenarios, the static spherically symmetric metric and Robertson-Walker metric, $g_{0i}$ vanishes. For our system, the energy-momentum tensor is
\begin{subequations}\label{eq:22}
\ba
\Too &=& \dot{\theta}^2 - \tfrac{1}{2}\ \gdoo\ \goo\, \dot{\theta}^2 = \tfrac{1}{2}\ \dot{\theta}^2 \label{eq:22a},\\
\Tij &=& -\tfrac{1}{2}\ g_{ij}\ \goo\, \dot{\theta}^2, \label{eq:22b}\\
\Toi &=& 0. \label{eq:22c}
\ea
\end{subequations}
On the other hand, for the ideal fluid the energy-momentum tensor is of the form
\be
\Ttmn = - P\, \gdmn + (P +\rho)\ u_\mu u_\nu,
\label{eq:23}
\ee
where $P$ is the pressure, $\rho$ is the energy density and $u^\mu$  is the four-velocity of the fluid normalized to unity, $u^\mu u^\nu\, \gdmn = 1$. In the comoving coordinates where the fluid is at rest, $u^\mu = (1/\sqrt{\gdoo}, \, {\bf 0})$, the energy momentum tensor is,
\begin{subequations}\label{eq:24}
\ba
\Ttoo &=& \gdoo\, \rho, \label{eq:24a}\\
\Ttij &=& - g_{ij}\, P, \label{eq:24b}\\
\Ttoi &=& 0. \label{eq:24c}
\ea
\end{subequations}
Comparison with (\ref{eq:22}) shows that
\be
\rho = P = \tfrac{1}{2}\ \goo\, \dot{\theta}^2. 
\label{eq:25}
\ee

A fluid with this equation of state is  called a ``stiff" fluid. For our two scenarios, we have
\be
\rho (r) = \frac{1}{16\pi\, G r^2} 
\label{eq:26} 
\ee
for the static, spherically symmetric space given in (\ref{eq:13}) and
\be
\rho (t) = \frac{1}{24\pi\, G t^2} 
\label{eq:27}
\ee
for the cosmological solution with $a(t) \sim t^{1/3}$. Note that (\ref{eq:26}) again shows the singular behavior at $r=0$.

The rather vast gravity-fluid mechanics literature, with an arbitrary equation of state, can be searched for our static solution, which corresponds to a ``stiff" fluid. The only ``stiff" fluid space-time that can be found in that literature is the Wyman solution. \cite{pi04}.

This work was supported by the U.S. Department of Energy (D.O.E) under the cooperative research agreement DE-FG02-05ER41360.


\end{document}